\documentclass{ws-mpla-hep}
 \usepackage{amsmath,amssymb}

\def\muF{\relax\ifmmode\mu_\text{F}^2\else{$\mu_\text{F}^2${ }}\fi}
\def\muR{\relax\ifmmode\mu_\text{R}^2\else{$\mu_\text{R}^2${ }}\fi}

\begin{document}

\def\preprint{RUB-TPII-11/09}

\markboth{N.~G.~Stefanis and I.~O.~Cherednikov}
{Renormalization-group anatomy of TMD PDFs in QCD}

\catchline{}{}{}{}{}

\title{\normalsize Renormalization-group anatomy of
transverse-momentum dependent parton distribution functions in QCD\\
      }

\author{\footnotesize{N.~G.~STEFANIS}\footnote{Talk presented at 
Workshop on ``Recent Advances in Perturbative QCD 
and Hadronic Physics'',
20--25 July 2009, ECT*, Trento, Italy, 
in Honor of Prof. Anatoly Efremov's 75th birthday.}
\; $\!\!$\footnote{\footnotesize{Also at
Bogoliubov Laboratory of Theoretical Physics,
JINR, 141980 Dubna, Russia.}}}

\address{\footnotesize{Institut f\"{u}r Theoretische Physik II,
Ruhr-Universit\"{a}t Bochum,
D-44780 Bochum, Germany}
stefanis@tp2.ruhr-uni-bochum.de}

\author{\footnotesize{I.~O.~CHEREDNIKOV}\footnote{\footnotesize{Also at
Institute for Theoretical Problems of Microphysics,
Moscow State University, 119899 Moscow, Russia.}}}

\address{\footnotesize{Bogoliubov Laboratory of Theoretical Physics,
JINR, 141980 Dubna, Russia \\
INFN Cosenza, Universit$\grave{a}$
della Calabria, I-87036 Rende (CS), Italy}\\
igor.cherednikov@jinr.ru}

\maketitle

\pub{Received (Day Month Year)}{Revised (Day Month Year)}

\begin{abstract}
The ultraviolet and rapidity divergences of transverse-momentum
dependent parton distribution functions with lightlike and transverse
gauge links is studied, also incorporating a soft eikonal factor.
We find that in the light-cone gauge with $q^-$-independent pole
prescriptions extra divergences appear which amount, at one-loop,
to a cusp-like anomalous dimension.
We show that such contributions are absent when the
Mandelstam-Leibbrandt prescription is used.
In the first case, the soft factor cancels the anomalous-dimension
defect, while in the second case its ultraviolet-divergent part
reduces to unity.

\end{abstract}

\ccode{PACS numbers: 11.10Jj, 13.85.Hd}

\section{Introduction}
\label{sec:intro}
In recent years, with the observation of observables related to
transversity, there has been renewed theoretical
attention\cite{Col08,Bacch08,CRS07}
to the gauge invariance of QCD operators in the definition of
unintegrated, i.e., transverse-momentum dependent (TMD), parton
distribution functions (PDF)s.\cite{CS81,CS82}
The core components of such quantities are gauge links (also called
Wilson lines) which are, in general, gauge-contour dependent and have,
therefore, more complicated renormalization properties ensuing from
their contour obstructions: end-points, cusps, and
self-intersections.\cite{Pol79}
In fact, as we pointed out in Refs.\ \refcite{CS07,CS08}, the
gauge-invariant definition of TMD PDFs,\cite{BJY02,JY02,BMP03},
which includes gauge links in the transverse direction, inevitably
involves gauge contours with a cusped-like junction point.
The renormalization effect on this point induces extra ultraviolet
(UV) divergences which give rise to an anomalous dimension that in
one-loop order coincides with the corresponding term of the
universal cusp anomalous dimension.\cite{KR87}

We argued in Refs.\ \refcite{CS07,CS08} that in order to counter this problem,
the fully gauge-invariant definition of the TMD PDFs has to include an
additional soft (eikonal) factor\cite{CH00,Hau07,CM04} along a special
cusped gauge contour (see Sec.\ \ref{sec:redefinition}) which extends
to the light-cone infinity in the transverse direction and serves to
compensate the spurious contribution to the anomalous dimension
peculiar to the light-cone gauge in connection with the advanced,
retarded, or principal-value prescription.
On the other hand, we have demonstrated\cite{CS09} that adopting
instead the Mandelstam-Leibbrandt (ML) pole
prescription\cite{Man83,Lei84} for the gluon propagator, such
contributions are absent so that the anomalous dimension of the
TMD PDF coincides with the result in covariant gauges, while the
UV-divergent part of the soft factor reduces to unity.

\section{Gauge links in TMD PDFs}
\label{sec:gauge-links}

The study of inclusive processes, e.g., deeply inelastic scattering
(DIS), involves integrated parton distribution functions which have the
following gauge-invariant definition
($i$ labels the sort of the parton in hadron $h$)
\begin{eqnarray}
  f_{i/h} \left(x, Q^2\right)
=
  \frac{1}{2}
  \int_{}^{} \frac{d\xi^{-}}{2\pi}
  {\rm e}^{- i k^{+} \xi^{-}}
  \langle h(P)
         |  \bar{\psi}_i(\xi^{-})
         \left[\xi^{-}, 0^{-}\right]
         \gamma^{+} \psi_{i}\left(0^{-}\right)
  |h(P) \rangle \Big|_{\xi^{+},\,
  \mbox{\footnotesize\boldmath$\scriptstyle\xi_\perp$} = 0}
\label{eq:ipdf}
\end{eqnarray}
and have renormalization properties controlled by the
Dokshitzer-Gribov-Lipatov-Altarelli-Parisi (DGLAP)
evolution equation.
Gauge invariance is ensured via the gauge link
\begin{equation}
  [\xi^-,0^-]
=
  \mathcal{P} \exp
  \left[
        -ig \int_{0^-}^{\xi^-} dz^\mu
        A^{a}_{\mu}(0^+,z^-,\mathbf{0}_\perp)\, t^a
  \right]
\label{eq:link}
\end{equation}
along a gauge contour on the light cone.

On the other hand, the study of semi-inclusive processes, such as
semi-inclusive deeply inelastic scattering (SIDIS), or the Drell-Yan
(DY) process---where one more final or initial hadron is detected and
its transverse momentum is observed---requires the introduction of
more complicated quantities, viz., distribution or fragmentation
functions depending on the parton's intrinsic transverse momentum
$\mathbf{k}_\perp$, which is, therefore, kept unintegrated.
In that case, the imposition of the light-cone gauge $A^+=0$ is
not sufficient to exhaust the gauge freedom in the TMD PDF.\cite{BJY02}
One can still perform $x^-$-independent gauge transformations
without changing the gauge condition.
The way out is to include into the definition of the TMD PDF
also a gauge link in the transverse direction off the light
cone.\cite{BJY02}
Then, in order to eliminate both gauge links, that along the light
cone and the transverse one, one has to adopt the light-cone (LC)
gauge $A^+=0$ together with the advanced boundary condition to go
around the poles of the gluon propagator in the complex
$\left(\mathfrak{Re}\, q^+, \mathfrak{Im}\, q^+\right)$ (or $q^0$)
plane.\cite{CRS07}
The gluon propagator in the $A^+=0$ gauge has the form
\begin{equation}
  D_{\mu\nu}^{\rm LC}(q)
=
  \frac{-i}{q^2-\lambda^2+i\epsilon}
  \left(g_{\mu\nu}
  - \frac{q_{\mu}n_{\nu}^{-} + q_{\nu}n_{\mu}^{-}}{[q^{+}]}
  \right)
\label{eq:gluon-prop}
\end{equation}
and has to be evaluated according to one of the following pole
prescriptions
\begin{equation}
  \frac{1}{[q^+]}\Bigg|_{\rm Ret/Adv}
=
  \frac{1}{q^+ \pm i \eta} \ , \ \ \
  \frac{1}{[q^+]}\Bigg|_{\rm PV}
=
  \frac{1}{2} \left[  \frac{1}{q^{+} + i \eta}
                + \frac{1}{q^{+} - i \eta}
              \right] \, ,
\label{eq:pole-prescr}
\end{equation}
where $\eta$ has the dimension of mass and is kept small but finite,
while the collinear poles are controlled by the quark virtuality
$p^2 < 0$, and the infrared (IR) singularities are regularized by the
auxiliary gluon mass $\lambda$.
None of these regulators will appear in the final expressions for
the anomalous dimensions.

Hence, one has the following operator definition
\begin{eqnarray}
  && f_{i/h} \left(x, \mbox{\boldmath$k_\perp$};\mu^2 \right)
=
  \frac{1}{2}
  \int \frac{d\xi^- d^2\mbox{\boldmath$\xi_\perp$}}{2\pi (2\pi)^2}
  {\rm e}^{-ik^{+}\xi^{-} +i \mbox{\footnotesize\boldmath$k_\perp$}
\cdot \mbox{\footnotesize\boldmath$\xi_\perp$}}
  \left\langle
              h |\bar \psi_i (\xi^-, \mbox{\boldmath$\xi_\perp$})
              [\xi^-, \mbox{\boldmath$\xi_\perp$};
   \infty^-, \mbox{\boldmath$\xi_\perp$}]^\dagger
\right. \nonumber \\
&& \left. \times
   [\infty^-, \mbox{\boldmath$\xi_\perp$};
   \infty^-, \mbox{\boldmath$\infty_\perp$}]^\dagger
   \gamma^+[\infty^-, \mbox{\boldmath$\infty_\perp$};
   \infty^-, \mbox{\boldmath$0_\perp$}]
   [\infty^-, \mbox{\boldmath$0_\perp$}; 0^-,\mbox{\boldmath$0_\perp$}]
   \psi_i (0^-,\mbox{\boldmath$0_\perp$}) | h
   \right\rangle \, ,
\label{eq:tmd_naive}
\end{eqnarray}
where gauge invariance is ensured by means of the path-ordered
gauge links
\begin{eqnarray}
\begin{split}
  [\infty^-, \mbox{\boldmath$z_\perp$}; z^-, \mbox{\boldmath$z_\perp$}]
\equiv {} &
 {\cal P} \exp \left[
                     i g \int_0^\infty d\tau \ n_{\mu}^- \
                      A_{a}^{\mu}t^{a} (z + n^- \tau)
               \right]~~~\mbox{lightlike link} \, , \\
 [\infty^-, \mbox{\boldmath$\infty_\perp$};
 \infty^-, \mbox{\boldmath$\xi_\perp$}]
\equiv {}&
 {\cal P} \exp \left[
                     i g \int_0^\infty d\tau \ \mbox{\boldmath$l$}
                     \cdot \mbox{\boldmath$A$}_{a} t^{a}
                     (\mbox{\boldmath$\xi_\perp$}
                     + \mbox{\boldmath$l$}\tau)
               \right] ~~~\mbox{transverse link} \, .
\end{split}
\end{eqnarray}
Note that the two-dimensional vector $\mbox{\boldmath $l$}$ is
arbitrary with no influence on the (local) anomalous dimensions
(i.e., it drops out from the final results).
Physically, the advanced prescription means that all final-state
gluon interactions between the struck quark and the spectators have
been reshuffled from the final to the initial state and have
been absorbed into the corresponding wave function.\cite{JY02,BJY02}
\begin{figure}[ph]
\centerline{\psfig{file=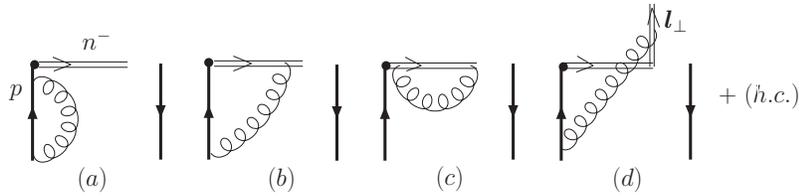,width=1.0in,angle=90,clip=}}
\vspace*{8pt}
\caption{One-loop gluon radiative corrections (curly lines)
contributing UV divergences to the TMD PDF with gauge links
(double lines).
Diagrams (a), (b), and (c) appear in covariant gauges, while
in light-cone gauges only diagrams (a) and (d) contribute.
The ``mirror'' (Hermitian conjugate) diagrams---abbreviated by
$(h.c.)$---are not shown.
\protect\label{fig:fig1}}
\end{figure}

The one-loop diagrams (see Fig.\ \ref{fig:fig1}) in the light-cone
gauge with the advanced prescription, contributing to the TMD
distribution of a quark in a quark, have been calculated in Refs.\
\refcite{BJY02,JMY04}.
There is a concealed assumption in these calculations.
This is that the junction point of the two transverse gauge links at
light-cone infinity is such that the two gauge (i.e., integration)
contours are joined smoothly.
However, we have shown\cite{CS07,CS08} by explicit calculation of the
UV divergences of these diagrams that extra contributions appear that
generate an anomalous dimension equal---at this loop order---to the
universal cusp anomalous dimension.\cite{KR87}
[Whether this finding remains valid in higher loop orders has not been
proved yet.]
This indicates that the transverse gauge links off the light cone
have a cusp-like junction point whose renormalization gives a
nontrivial contribution to the anomalous dimension of the TMD PDF.

\section{One-loop radiative corrections of TMD PDFs}
\label{sec:rad-corr}

\subsection{$q^-$-independent pole prescriptions}
\label{sec:ML-pres}

Let us now look at the diagrams in Fig.\ \ref{fig:fig1} in more
detail.
First some general remarks.
Diagrams (a), (b), and (c) contribute in covariant gauges,
whereas in the light-cone gauge only diagrams (a) and (d) contribute.
Diagram (d) is associated with the transverse gauge link and, hence,
it is peculiar to the light-cone gauge.
Note that the Hermitian-conjugate diagrams (labeled $h.c.$) are
generated by ``mirror'' diagrams which are not shown explicitly.

First, we consider the imposition of $q^-$-independent pole
prescriptions, like those in Eq.\ (\ref{eq:pole-prescr}).
In that case, the transverse component of the gauge field reads
\begin{equation}
   \mathbf{A}^\perp (\infty^-; \mbox{\boldmath$\xi$}_\perp)
=
   \frac{g}{4\pi} C_\infty
   \mbox{\boldmath $\nabla$}^\perp
   \ln \Lambda |\mbox{\boldmath$\xi$}_\perp| \, ,
\label{eq:ret}
\end{equation}
where the numerical constant $C_\infty$ depends on the pole
prescription to treat the light-cone divergences in the notation of
Ref.\ \refcite{BJY02}
\begin{equation}
  C_\infty
=
  \left\{
  \begin{array}{ll}
  & \ \ 0  \ , \ {\rm Advanced}    \\
  & - 1 \ , \ {\rm Retarded}  \\
  & - \frac{1}{2} \ , \ {\rm Principal~Value~}\, ,
  \
  \end{array} \right.
\label{eq:c_inf}
\end{equation}
and $\Lambda$ is an IR regulator that does not enter the final
results.

Using dimensional regularization within the
$\overline{\text{MS}\vphantom{^1}}$ scheme, one can now compute
the contributions of diagrams (a) and (d) in Fig.\ \ref{fig:fig1}:
$\Sigma^{(a)}$ and $\Sigma^{(d)}$.
Diagram (a) contains UV divergences which translate into a single
pole $1/\epsilon$.
In addition, it contains divergences which are related to the
poles of the gluon propagator in the $A^+=0$ gauge and are controlled
by the employed pole prescription---embodied in the factor $C_\infty$.
Moreover, it also contains a UV-finite (i.e., $\epsilon$-independent)
part, which has, however, rapidity divergences---i.e., a term
proportional to $C_\infty$.
The explicit expressions for all these terms can be found in Ref.\
\refcite{CS08}.

On the other hand, diagram (d), which, we repeat, is absent in
covariant gauges, contains rapidity divergences driven by $C_\infty$
(see in Ref.\ \refcite{CS08} for the explicit expression and further
details).
It turns out that---including the mirror diagrams of (a) and (d)---this
contribution exactly cancels all terms proportional to
$C_\infty$ in $\Sigma^{(a)}_{\rm UV}$ and also in
$\Sigma^{(a)}_{\rm finite}$, rendering the total expression
pole-prescription independent.
What remains reads
\begin{eqnarray}
  {\Sigma}^{(a+d)}_{\rm UV} (p, \mu, \alpha_s ; \epsilon)
& = &
  - \frac{\alpha_s}{\pi}\ C_{\rm F} \frac{1}{\epsilon}
    \left[
          \frac{1}{4}- \frac{\gamma^{+} \hat p}{2 p^{+}}
    \left(1 +  \ln \frac{\eta}{p^{+}} - \frac{i\pi}{2}
          - i \pi \ C_{\infty}^{(a)} + i \pi C_{\infty}^{(d)}
  \right)
    \right]
\nonumber \\
& = &
  - \frac{\alpha_s}{\pi}\ C_{\rm F} \frac{1}{\epsilon}
    \left[1 - \frac{\gamma^{+} \hat p}{2 p^{+}}
    \left(1 + \ln \frac{\eta}{p^{+}} - \frac{i\pi}{2} \right)
    \right]
    \, ,
\label{eq:s_tot}
\end{eqnarray}
where
$C_{\rm F}=\left(N_{\rm c}^{2}-1\right)/2N_{\rm c}=4/3$
and the superscripts $(a)$ and $(d)$ on $C_\infty$ serve to
indicate the diagram from which the corresponding contribution
originates.
The above expression can be further evaluated by virtue of
$
  \frac{ \gamma^+ \hat p \gamma^+}{2 p^+} = \gamma^+
$
to obtain
\begin{equation}
  \Sigma_{\rm UV}^{\rm (a+d)}(\alpha_s, \epsilon)
=
   2\frac{\alpha_s}{\pi}C_{\rm F} \left[ \frac{1}{\epsilon}
   \left( \frac{3}{4}
  + \ln \frac{\eta}{p^+} \right) - \gamma_E + \ln 4\pi \right] \, .
\label{eq:sigma_a+d}
\end{equation}
From this expression we find
\begin{eqnarray}
  \gamma_{\rm LC}
 =
  \frac{\mu}{2}
  \frac{1}{Z}
  \frac{\partial\alpha_s}{\partial\mu}
  \frac{\partial Z}{\partial\alpha_s}
=
  \frac{\alpha_s}{\pi}C_{\rm F}\Bigg( \frac{3}{4}
  + \ln \frac{\eta}{p^+} \Bigg)
=
  \gamma_{\rm smooth} - \delta \gamma \, .
\label{eq:gamma_LC}
\end{eqnarray}
Thus, the one-loop anomalous dimension in the LC gauge with
$q^-$-independent pole prescriptions reads
\begin{equation}
  \gamma_{\rm LC}
=
  \gamma_{\rm smooth} - \delta \gamma , \ \ \  \gamma_{\rm smooth}
=
  \frac{3}{4} \frac{\alpha_s}{\pi}C_{\rm F} + O(\alpha_{s}^{2}) \ , \ \ \
  \delta \gamma
=
  - \frac{\alpha_s}{\pi}C_{\rm F} \ln \frac{\eta}{p^+} \, ,
\end{equation}
where the defect of the anomalous dimension $\delta \gamma$
encapsulates the deviation of the calculated quantity from the
anomalous dimension of the two-quark operator with a
gauge-connector insertion in a covariant gauge.\cite{CD80,Ste83}
Note that the only anomalous dimensions ensuing from the gauge
connector along any smooth contour stem from its endpoints.
Hence, the backfit logic in this finding is to modify the definition
of the TMD PDF in Eq.\ (\ref{eq:tmd_naive}) in such a way as to
dispense with $\delta\gamma$.
But to do so, we have first to understand the deeper meaning of
the anomalous-dimension defect.
To this end, write $p^+ = (p \cdot n^-) \sim \cosh \chi$ and observe
that it defines an angle $\chi$ between the direction of the quark
momentum $p_\mu$ and the lightlike vector $n^-$.
In the large $\chi$ limit, $\ln p^+ \to \chi , \ \chi \to \infty$.
Therefore, it appears natural to conclude that the defect of the
anomalous dimension $\delta \gamma$ can be identified with the
universal cusp anomalous dimension at the one-loop order:\cite{KR87}
\begin{equation}
\begin{split}
   & \gamma_{\rm cusp} (\alpha_s, \chi)
= \frac{\alpha_s}{\pi}C_{\rm F} \ (\chi \coth \chi - 1 ) \, , \\
& \frac{d}{d \ln p^+} \ \delta \gamma
= \lim_{\chi \to \infty}
  \frac{d}{d \chi} \gamma_{\rm cusp} (\alpha_s, \chi)
= \frac{\alpha_s}{\pi}C_{\rm F} \, .
\label{eq:cusp-an-dim}
\end{split}
\end{equation}

\subsection{Mandelstam-Leibbrandt pole prescription}
\label{subsec:ML-pres}

\begin{figure}[h]
\centerline{\psfig{file=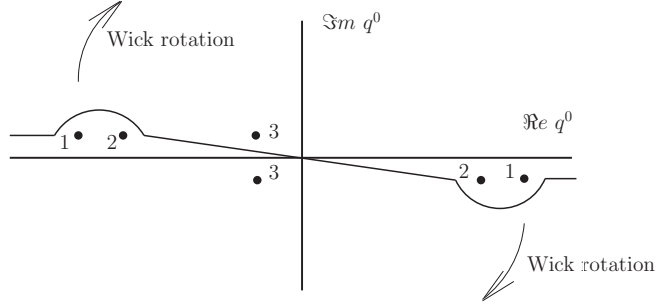,width=1.6in,angle=90,clip=}}
\vspace*{8pt}
\caption{Integration contour and poles of the gluon propagator in the
         $\left(\mathfrak{Re}\, q^0, \mathfrak{Im}\, q^0\right)$
         plane. The results of the ML-prescription (position 1) and
         those in a covariant gauge (position 2) belong to the same
         quadrants: second and fourth---not valid for the
         principal-value prescription (position 3).
\protect\label{fig:fig2}}
\end{figure}

Let us now consider the application of the light-cone gauge in
conjunction with a $q^-$-dependent pole prescription for the gluon
propagator.
Imposing the Mandelstam-Leibbrandt (ML) pole
prescription,\cite{Man83,Lei84} which depends on both variables
$q^+$ and $q^-$, gives rise to a more complicated structure of the
gluon propagator in the complex $q^0$ plane, viz.,
\begin{equation}
  \frac{1}{[q^+]_{\rm ML}}
= \left\{
  \begin{array}{ll}
  &  \frac{1}{q^+ + i0q^-}     \\
  &  \frac{q^-}{q^+q^- + i0}
  \
  \end{array} \right.
  \, ,
\label{eq:MLdef}
\end{equation}
which is illustrated graphically in Fig.\ \ref{fig:fig2}.

Our goal is to evaluate diagrams $(a)$ and $(d)$ in Fig.\
\ref{fig:fig1} using this pole prescription.
In that case, the evaluation of $\Sigma^{(a)}$ is more complicated and
we refer for the technical steps to Ref.\ \refcite{CS09}.
Here we only sketch the main results.
First, recall that $\Sigma^{(a)}$ contains a part proportional to
$g_{\mu\nu}$---Feynman term---which is pole-prescription independent.
The rest, labeled $\Sigma^{(a)}_{\rm ML}$, can be computed analytically
to yield
\begin{eqnarray}
  {\Sigma}^{(a)} (p, \alpha_s ; \mu, \epsilon)
& = &
  {\Sigma}^{(a)}_{\rm Feynman} + {\Sigma}^{(a)}_{\rm ML}
\nonumber \\
& = &
  - \frac{\alpha_s}{4\pi} C_{\rm F} \  \Gamma(\epsilon)
    \left(-4\pi\frac{\mu^2}{p^2}\right)^\epsilon
    \frac{\Gamma^2(1-\epsilon)}{\Gamma(2 - 2\epsilon)}
    \left[ (1-\epsilon) - 4\right] \, .
\label{eq:sigma-40}
\end{eqnarray}
Extracting the UV divergent terms in the $\overline{\rm MS}$-scheme,
one gets (after adding the mirror diagrams):
\begin{equation}
  {\Sigma}_{\rm UV/ML}^{(a)} (p, \alpha_s,  \mu , \epsilon )
=
  - \frac{\alpha_s}{4\pi} C_{\rm F}
  \left[\frac{1}{\epsilon}(1 - 4) - \gamma_E + 4\pi\right]
=
  - \frac{3 \alpha_s}{4\pi} C_{\rm F}
  \left[ \frac{1}{\epsilon} - \gamma_E + 4\pi\right] \, .
\label{eq:fey4}
\end{equation}
Obviously, employing the ML-pole prescription in the light-cone
gauge, both the UV-divergent part of the TMD PDF and also the finite
one do not contain any extra terms of the form $\ln p^+$ which
could be attributed to a cusped contour---in contrast to the
findings of the previous subsection.

The next task is to evaluate diagram $(d)$, which contains the
cross-talk of the transverse gauge links
$
 \left[
       \infty^-, 0^+; \mbox{\boldmath $l$}_{\perp} \tau
 \right] \, ,
 \left[
       \infty^-, 0^+; \mbox{\boldmath $l$}_{\perp} \tau
       + \mbox{\boldmath $\xi$}_{\perp}
 \right]
$.
To evaluate this expression, we have first to determine the transverse
gauge field at light-cone infinity.
This calculation was performed in Ref.\ \refcite{CS09}, from which we
quote the result
\begin{equation}
  \mathbf{A}^\perp (\infty^-; \mbox{\boldmath$\xi$}_\perp)
=
  - \frac{g}{4\pi}
  \mbox{\boldmath $\nabla$}^\perp \ln
  \Lambda |\mbox{\boldmath$\xi$}_\perp| \, .
\label{eq:MLret}
\end{equation}
Then, we find
\begin{eqnarray}
   {\Sigma}_{\rm ML}^{(d)} (p, \mu , g; \epsilon)
=
   - g^2 C_{\rm F} \mu^{2\epsilon} 2\pi i  \int\!
   \frac{d^\omega q}{(2\pi)^\omega} \delta (q^+) \
   \frac{\gamma^+ (\hat p - \hat q)}{(p-q)^2 \ q^2} \, .
\label{eq:perp}
\end{eqnarray}
Collecting the UV divergences of both diagrams $(a)$ and $(d)$, we
obtain
\begin{equation}
  {\Sigma}^{(a+d)}_{\rm ML/UV} (p, \mu, \alpha_s ; \epsilon)
=
  - \frac{\alpha_s}{\pi}\ C_{\rm F}
  \Bigg\{ \frac{1}{\epsilon}
                            \left[\frac{1}{4} - \frac{\gamma^+ \hat p}
                                                     {2 p^+}
                                                \left( 1-\frac{i\pi}{2}
                                                \right)
                            \right] - \gamma_E + 4 \pi
  \Bigg\} \, ,
\label{eq:s_tot_ML}
\end{equation}
which finally yields
($
  \gamma^{+} \hat p \gamma^{+}/2 p^{+} = \gamma^{+}
$)
\begin{equation}
  {\Sigma}^{(a+d)}_{\rm ML/UV} (p, \mu, \alpha_s ; \epsilon)
=
  \frac{\alpha_s}{\pi}\ C_{\rm F}
                                 \left[ \frac{1}{\epsilon}
                                 \left(\frac{3}{4} + \frac{i\pi}{2}
                                 \right)
                                 - \gamma_E + 4\pi
                                 \right] \, .
\label{eq:s_tot-final_ML}
\end{equation}
After including the mirror contribution to graph $(d)$ in Fig.\
\ref{fig:fig1}, one arrives at the following expression
\begin{equation}
  {\Sigma}^{(a+d)}_{\rm ML/UV} (p, \mu, \alpha_s ; \epsilon)
=
  \frac{\alpha_s}{\pi} C_{\rm F}
                               \left[ \frac{1}{\epsilon}
                                      \frac{3}{4} - \gamma_E + 4\pi
                               \right]\, ,
\label{eq:s_tot-final}
\end{equation}
which does not contain an imaginary part and resembles the result
one finds in covariant gauges.

Bottom line: Using the Mandelstam-Leibbrandt pole prescription to treat
the rapidity divergences in the gluon propagator in association with
the light-cone gauge, no anomalous-dimension defect appears, so that
one gets the well-known expression
\begin{equation}
  \gamma_{\rm ML}
  =
  - \frac{1}{2} \mu \frac{d}{d\mu}
  \ln \Sigma_{\rm ML/UV}^{(a+d)} (\alpha_s, \epsilon)
=
  \frac{3}{4} \frac{\alpha_s}{\pi}\, C_{\rm F} + O(\alpha_{s}^{2})\, .
\end{equation}

\section{Soft factor and TMD PDF redefinition}
\label{sec:redefinition}

\begin{figure}[h]
\centerline{\psfig{file=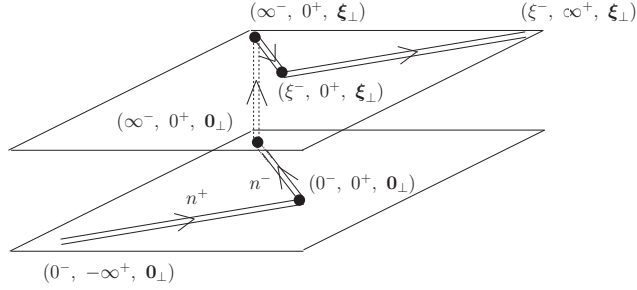,width=3.3in}}
\vspace*{8pt}
\caption{Cusped gauge contour associated with the soft factor $R$.
\protect\label{fig:fig3}}
\end{figure}

Aiming for a more suitable definition of TMD PDFs, we propose to
refurbish Eq.\ (\ref{eq:tmd_naive}) by including a soft factor $R$
\begin{equation}
  R
\equiv
 \left\langle 0
 \left| {\cal P}
 \exp\Bigg[ig \int_{\Gamma_{\rm cusp}}\!\!\!\!\! d\zeta^\mu
 \ t^a A^a_\mu (\zeta)\Big] \cdot
 {\cal P}^{-1} \exp\Big[-ig \int_{\Gamma_{\rm cusp}'}d\zeta^\mu
 \ t^a A^a_\mu (\xi + \zeta)\Bigg]
 \right|0
 \right\rangle \, , ~~~
\label{eq:soft_factor_1}
\end{equation}
that contains eikonal lines, giving rise to an anomalous dimension which
is equal in magnitude but opposite in sign to the defect of the
anomalous dimension entailed by the cusped junction point of the gauge
contours.
This soft factor is calculated along the particular gauge contour
shown in Fig.\ \ref{fig:fig3}.

The involved gauge contours are defined as follows:
\begin{eqnarray}
&& \nonumber \Gamma_{\rm cusp} : \zeta_\mu
=
  \{ [p_\mu^{+}s, \ - \infty < s < 0]
 \cup [n_\mu^-  s' ,
  \ 0 < s' < \infty] \cup
  [ \mbox{\boldmath$l_\perp$} \tau , 0 < \tau < \infty ] \}
\\
&& \Gamma_{\rm cusp}' : \zeta_\mu
=
  \{ [p_\mu^{+}s ,  \ + \infty < s < 0]
 \cup [n_\mu^-  s' ,
  \ 0 < s' < \infty] \cup
  [ \mbox{\boldmath$l_\perp$} \tau ,   0 < \tau < \infty ] \}\, .
~~~~~
\label{eq:gpm}
\end{eqnarray}
Then, the TMD PDF can be redefined to read
\begin{eqnarray}
  f_{i/h}^{\rm mod}\left(x, \mbox{\boldmath$k_\perp$};\mu^2\right)
&& \!\!\! =
  \frac{1}{2}
  \int \frac{d\xi^- d^2\mbox{\boldmath$\xi_\perp$}}{2\pi (2\pi)^2}
  {\rm e}^{-ik^+\xi^- +i { \mbox{\footnotesize\boldmath$k_\perp$}}
  \cdot \mbox{\footnotesize\boldmath$\xi_\perp$}}
  \left\langle
              h |\bar \psi_i (\xi^-, \mbox{\boldmath$\xi_\perp$})
              [\xi^-, \mbox{\boldmath$\xi_\perp$};
   \infty^-, \mbox{\boldmath$\xi_\perp$}]^\dagger
\right. \nonumber \\
\rule{0in}{3ex}
&& \left. \times
   [\infty^-, \mbox{\boldmath$\xi_\perp$};
   \infty^-, \mbox{\boldmath$\infty_\perp$}]^\dagger
   \gamma^+[\infty^-, \mbox{\boldmath$\infty_\perp$};
   \infty^-, \mbox{\boldmath$0_\perp$}]
   [\infty^-, \mbox{\boldmath$0_\perp$}; 0^-,\mbox{\boldmath$0_\perp$}]
\right. \nonumber \\
\rule{0in}{4ex}
&& \left. \times
   \psi_i (0^-,\mbox{\boldmath$0_\perp$}) |h
   \right\rangle
   R (p^+, n^-|\xi^{-}, \mbox{\boldmath$\xi_\perp$}) \ \, .
\label{eq:tmd_re-definition}
\end{eqnarray}

The one-loop radiative corrections to $R$ originate from diagrams
analogous to those shown in Fig.\ \ref{fig:fig1} in which the
thick quark line is replaced by a double line associated with the
gauge link entering Eq.\ (\ref{eq:soft_factor_1}).
Employing the light-cone gauge $A^+=0$ with one of the $q^-$-independent
pole prescriptions [Adv, Ret, PV] according to
Eq.\ (\ref{eq:pole-prescr}), we get for the UV part the following
expression
\begin{equation}
  \Phi^{\rm (a + d)}_{\rm UV/LC}(\epsilon; \eta)
=
  - \frac{\alpha_s}{\pi}C_{\rm F}\frac{1}{\epsilon}
  \left(
        \ln \frac{\eta}{p^+} - i\frac{\pi}{2}
  \right) \, .
\label{eq:soft_ad}
\end{equation}
Taking into account the mirror diagrams to $(a)$ and $(d)$ (with the
thick quark line replaced by a double line for the gauge link), one
obtains the total UV-divergent part of the soft factor at the one-loop
order:
\begin{equation}
  \Phi^{\rm 1-loop}_{\rm UV/LC}(\epsilon; \eta)
=
  - \frac{\alpha_s}{\pi}C_{\rm F}  \frac{2}{\epsilon}
    \ln \frac{\eta}{p^+} \, .
\label{eq:phase_1loop}
\end{equation}
One notes that this expression bears no dependence on the pole
prescription, i.e., all $C_\infty$-dependent terms are absent.
The only surviving contribution to the associated anomalous dimension
is the cusp-related term $\sim \ln p^+$ which amounts to
$-\gamma_{\rm cusp}$.
This completes the proof that the modified TMD PDF definition
(\ref{eq:tmd_re-definition}) contains no artifacts stemming from
contour obstructions.

Consider now the case with the ML pole prescription.
As we have already discussed, this pole prescription removes all
undesirable light-cone singularities.
Therefore, we should expect that the insertion of the soft factor $R$
in that case does not destroy this property.
Skipping details, let us remark that the UV singularities of the soft
factor are generated by the self-energy of the lightlike gauge link
and the one-gluon exchanges between the lightlike and the transverse
gauge link (analogous diagrams to $(a)$ and $(d)$ in Fig.\
\ref{fig:fig1}).
Thus, one has
\begin{equation}
  \Phi_{\rm soft}^{\rm LO/ML}
=
  \Phi_{\rm soft}^{(0)} + \Phi_{\rm soft}^{(1)} + O(\alpha_s^2) \, ,
\end{equation}
with
$
  \Phi_{\rm soft}^{(0)} = 1
$
and
\begin{eqnarray}
  \ \ \Phi_{\rm soft}^{(1)}
& = &
  \Phi_{\rm soft-virt}^{(1)} + \Phi_{\rm soft-real}^{(1)} \nonumber \\
  \Phi_{\rm soft-virt}^{(1)}
& = &
  \Phi_{\rm soft-virt}^{(a)} + \Phi_{\rm soft-virt}^{(d)} \, ,
\end{eqnarray}
where
\begin{equation}
  \Phi_{\rm soft-virt}^{(a)}
=
  2 i g^2 \mu^{2\epsilon} C_{\rm F}\!
  \int_0^\infty d\sigma \int_0^\sigma d\tau \int\!\!
  \frac{d^\omega q}{(2\pi)^\omega}
  \frac{{\rm e}^{-i q^- (\sigma-\tau)}}{2 q^+q^-
  - \mbox{\boldmath $q$}_\perp^2 + i0}
  \frac{q^-}{q^+ + i0 q^-}
\end{equation}
and the vector $u_\mu$ is being chosen to be lightlike:
$
 u_\mu = (p^+, 0^-, \mbox{\boldmath $0$}_\perp)
$.
Appealing to Fig.\ \ref{fig:fig2}, we observe that the integral above
vanishes by virtue of the location of the poles in the Feynman and
ML denominators on the same side of the $q^+$-axis.
Consequently, we have
\begin{equation}
  \Phi_{\rm soft-virt}^{(a)\rm ML}
=
  0 \, .
\end{equation}
For the same reason, also the contribution of diagram $(d)$
vanishes, entailing $\Phi_{\rm soft-virt}^{(d)\rm ML} = 0$.
On the other hand, the contribution arising from real gluons,
$\Phi_{\rm soft-real}^{(1)}$, does not contain UV-singularities.
Hence, the UV-divergent part of $\Phi_{\rm soft}^{\rm LO/ML}$
reduces to unity, validating Eq.\ (\ref{eq:tmd_re-definition})
also for the case of the light-cone gauge with the ML-prescription.

\section{Evolution of the TMD PDF}
\label{sec:evol}

The evolution behavior of TMD PDFs is of particular theoretical and
phenomenological importance.\cite{CT05}
Theoretically, we have to verify whether the integrated PDF, obtained
from Eq.\ (\ref{eq:tmd_re-definition}), coincides with the standard one
with no contour artifacts left over.
Furthermore, we have to deal with the additional dependence on the
scale $\eta$ by means of an additional evolution equation.
Note that in our approach this mass parameter plays a role akin to the
rapidity parameter $\zeta$ in the Collins-Soper evolution
equation.\cite{CS81}
We have discussed\cite{CS08} some of these issues and shown that the
redefined TMD PDF satisfies the following renormalization-group
equation
\begin{equation}
  \frac{1}{2} \mu \frac{d}{d\mu}
  \ln f_{q/q}^{\rm mod}(x, \mbox{\boldmath$k_\perp$}; \mu, \eta)
=
  \frac{3}{4} \frac{\alpha_s}{\pi}\, C_{\rm F} + O(\alpha_s^2)
=
  \gamma_{f_{q/q}}
  \, .
\end{equation}
Taking logarithmic derivatives of
$f_{q/q}^{\rm mod}(x, \mbox{\boldmath$k_\perp$}; \mu, \eta)$
with respect to both scales $\mu$ and $\eta$, we get
\begin{equation}
  \mu \frac{d}{d\mu}
  \left[ \eta \frac{d}{d\eta}
         f_{q/q}^{\rm mod}(x, \mbox{\boldmath$k_\perp$}; \mu, \eta)
  \right]
=
  0 \, ,
\label{eq:cons}
\end{equation}
which establishes the formal analogy between our
approach and the Collins-Soper one.\cite{CS81}
We emphasize that only the modified definition via
Eq.\ (\ref{eq:tmd_re-definition}) satisfies such simple evolution
equations.

\section{Concluding Remarks}
\label{sec:concl}

We have discussed an approach to TMD PDFs which takes into account the
renormalization properties of the contour-dependent gauge links in
terms of the anomalous dimensions ensuing from contour obstructions.
We argued that supplementing the light-cone gauge with
$q^-$-independent pole prescriptions for the gluon propagator, leads to
the appearance of an anomalous dimension that can be associated with a
cusped junction point of the transverse gauge contours.
In contrast, we found that when the ML prescription is employed, which
depends on both variables $q^+$ and $q^-$, the defect of the anomalous
dimension cancels out.
In the first case, a nontrivial soft factor in the definition of the
TMD PDF restores the correct anomalous dimension by compensating the
renormalization effect on the cusp-like junction point of the contours.
In the second case, the soft factor reduces to unity and one recovers
the same result as in covariant gauges in which the transverse gauge
links are absent.

\section*{Acknowledgments}
We express our best wishes to A.\ V.\ Efremov on the occasion
of his 75th birthday.
This work was supported in part by the Heisenberg-Landau Program 2009
and the Russian Federation Scientific Schools Grant 195.2008.9.


\begin{thebibliography}{0}

\bibitem{Col08}
J.~Collins,
{\it Proc.\ Sci.} {\bf LC2008}, 028 (2008)
[arXiv:0808.2665 [hep-ph]].

\bibitem{Bacch08}
A.~Bacchetta, D.~Boer, M.~Diehl, and P.~J.~Mulders,
{\it JHEP} {\bf 0808}, 023 (2008).

\bibitem{CRS07}
J.~C.~Collins, T.~C.~Rogers, and A.~M.~Stasto,
{\it Phys.\ Rev.\ D} {\bf 77}, 085009 (2008).

\bibitem{CS81}
J.~C.~Collins and D.~E.~Soper,
{\it Nucl.\ Phys.\ B} {\bf 193}, 381 (1981);
{\it Nucl.\ Phys.\ B} {\bf 213}, 545 (1983) Erratum.

\bibitem{CS82}
J.~C.~Collins and D.~E.~Soper,
{\it Nucl.\ Phys.\ B} {\bf 194}, 445 (1982).

\bibitem{Pol79}
A.~M.~Polyakov,
{\it Nucl.\ Phys.\ B} {\bf 164}, 171 (1979).

\bibitem{CS07}
I.~O.~Cherednikov and N.~G.~Stefanis,
{\it Phys.\ Rev.\ D} {\bf 77}, 094001 (2008).

\bibitem{CS08}
I.~O.~Cherednikov and N.~G.~Stefanis,
{\it Nucl.\ Phys.\ B} {\bf 802}, 146 (2008).

\bibitem{BJY02}
A.~V.~Belitsky, X.~Ji, and F.~Yuan,
{\it Nucl.\ Phys.\ B} {\bf 656}, 165 (2003).

\bibitem{JY02}
X.~Ji and F.~Yuan,
{\it Phys.\ Lett.\ B} {\bf 543}, 66 (2002).

\bibitem{BMP03}
D.~Boer, P.~J.~Mulders, and F.~Pijlman,
{\it Nucl.\ Phys.\ B} {\bf 667}, 201 (2003).

\bibitem{KR87}
G.~P.~Korchemsky and A.~V.~Radyushkin,
{\it Nucl.\ Phys.\ B} {\bf 283}, 342 (1987).

\bibitem{CH00}
J.~C.~Collins and F.~Hautmann,
{\it Phys.\ Lett.\ B} {\bf 472}, 129 (2000).

\bibitem{Hau07}
F.~Hautmann,
{\it Phys.\ Lett.\ B} {\bf 655}, 26 (2007).

\bibitem{CM04}
J.~C.~Collins and A.~Metz,
{\it Phys.\ Rev.\ Lett.} {\bf 93}, 25001 (2004).

\bibitem{CS09}
I.~O.~Cherednikov and N.~G.~Stefanis,
{\it Phys.\ Rev.\ D} {\bf 80}, 054008 (2009).

\bibitem{Man83}
S.~Mandelstam,
{\it Nucl.\ Phys.\ B} {\bf 213}, 149 (1983).

\bibitem{Lei84}
G.~Leibbrandt,
{\it Phys.\ Rev.\ D} {\bf 29}, 1699 (1984).

\bibitem{JMY04}
X.~Ji, J.~Ma, and F.~Yuan,
{\it Phys.\ Rev.\ D} {\bf 71}, 034005 (2005).

\bibitem{CD80}
N.~S.~Craigie and H.~Dorn,
{\it Nucl.\ Phys.\ B} {\bf 185}, 204 (1981).

\bibitem{Ste83}
N.~G.~Stefanis,
{\it Nuovo Cim.\ A} {\bf 83}, 205 (1984).

\bibitem{CT05}
F.~A.~Ceccopieri and L.~Trentadue,
{\it Phys.\ Lett.\  B} {\bf 636}, 310 (2006);
%
{\it Phys.\ Lett.\  B} {\bf 660}, 43 (2008).

\end{thebibliography}
\end{document}